\documentclass[12pt]{article}

\usepackage{aaspp4,natbib,tighten,psfig}

\begin{document}


\def\ifundefined#1{\expandafter\ifx\csname#1\endcsname\relax}
\def\la{\mathrel{\hbox{\rlap{\hbox{\lower4pt\hbox{$\sim$}}}\hbox{$<$}}}}
\def\ga{\mathrel{\hbox{\rlap{\hbox{\lower4pt\hbox{$\sim$}}}\hbox{$>$}}}}
\ifundefined{nuc}\def\nuc#1#2{\relax\ifmmode{}^{#1}{\protect\text{#2}}\else${}^{#1}$#2\fi}\else\relax\fi
\ifundefined{ensuremath}\def\ensuremath#1{\relax\ifmmode{#1}}
\else\hbox{${#1}$}\fi\else\relax\fi
\def\doublespace{\setlength{\baselineskip}{23pt}}
\def\singlespace{\setlength{\baselineskip}{14pt}}
\newcommand{\Tmod}{\ensuremath{T_{\rm model}}}
 
%
%
\def\valid{}    

\def\la{\mathrel{\hbox{\rlap{\hbox{\lower4pt\hbox{$\sim$}}}\hbox{$<$}}}}
\def\ga{\mathrel{\hbox{\rlap{\hbox{\lower4pt\hbox{$\sim$}}}\hbox{$>$}}}}

\newcommand{\be}{\begin{eqnarray}}
\newcommand{\ee}{\end{eqnarray}}

\newcommand{\gcm}{g~cm$^{-3}$}
\newcommand{\is}{s$^{-1}$}
\newcommand{\msol}{\ifmmode{{\rm M}_\odot}\else{M$_\odot$}\fi}
\newcommand{\foe}{\ifmmode{10^{51}}\else{$10^{51}$}\fi}
\newcommand{\nni}{\nuc{56}{Ni}}
\newcommand{\xni}{\ifmmode{{\rm X}_{\rm Ni}}\else{X$_{\rm Ni}$}\fi}
\def\ang{\hbox{\AA}}
\def\Teff{\ifmmode{T_{\rm eff}}\else{\hbox{$T_{\rm eff}$} }\fi}
\def\Rzero{\ifmmode{R_0}\else{\hbox{$R_0$} }\fi}
\newcommand{\vno}{v_0}

\def\IBMSP{{\tt IBM} {\tt SP2}}
\def\Origin{{\tt SGI} {\tt Origin~2000}}
\def\HPJ{{\tt HP J200}}
\def\SP2{{\tt IBM SP2}}
\def\GCel{{\tt GCel}}
\def\GCpp{{\tt GC/PP}}
\def\MPI{{\tt MPI}}
\def\HPF{{\tt HPF}}
\def\PC2{{\tt PC$^2$}}
\def\piofs{{\tt PIOFS}}
\def\RT{radiative transfer}
\def\RTC{radiative transfer code}
\def\nrt{N_{\rm RT}}
\def\nnode{N_{\rm node}}
\def\logg{\log(g)}
\def\mh{[{\rm M/H}]}

\def\inu{\ifmmode{I_{\nu}}\else{\hbox{$I_{\nu}$} }\fi}
\def\snu{\ifmmode{S_{\nu}}\else{\hbox{$S_{\nu}$} }\fi}
\def\jnu{\ifmmode{J_{\nu}}\else{\hbox{$J_{\nu}$} }\fi}

\def\etal{et al.}
\def\fep{\ifmmode{{\rm Fe II}}\else\hbox{Fe~II }\fi}

\font\caps=cmcsc10                  
\font\dunh=cmdunh10  at 12.0 true pt 
\font\dunhs=cmdunh10 
\font\vbold=cmbx10 scaled \magstep1 
\font\sevenbf=cmbx7
\font\sevenit=cmti7
\font\Kapi=cmr17

\def\MEV{DOME}
\def\RTE{equation of radiative transfer}
\def\etal{{et al}}
\def\HW{H\&W}
\def\OK{O\&K}
\def\ok{O\&K}
\def\RH{R\&H}
\def\atlasIX{{\tt ATLAS9}}
\def\atlasXII{{\tt ATLAS12}}
\def\phoenix{{\tt PHOENIX}}
\def\PHOENIX{{\tt PHOENIX}}
\def\phx{{\tt PHOENIX}}
\def\IUE{{\bf IUE}}
\def\water{{H$_2$O}}

\def\MEV{DOME}
\def\RTE{equation of radiative transfer}
\def\etal{{et al}}
\def\HW{H\&W}
\def\OK{O\&K}
\def\ok{O\&K}
\def\RH{R\&H}
\def\phoenix{{\tt PHOENIX}}
\def\phx{{\tt PHOENIX}}
\def\IUE{{\bf IUE}}
\def\water{{H$_2$O}}
\def\jtio{J-TiO}
\def\jwater{J-\water}
\def\mtwater{MT-\water}
\def\hitran{{\tt HITRAN92}}

\def\ibmrs{\hbox{\tt RS/6000}}
\def\hp{\hbox{\tt HP~9000}}
\def\dec{\hbox{\tt DEC~5000}}
\def\axp{\hbox{\tt AXP}}
\def\ibmmf{\hbox{\tt IBM~3090}}
\def\ibmpc{\hbox{\tt 486DX}}
\def\cray{\hbox{\tt Cray 2}}
\def\ymp{\hbox{\tt YMP}}
\def\nec{\hbox{\tt NEC}}

\def\g{\gamma}
\def\b{\beta}
\def\m{\mu}
\def\e{\epsilon}
\def\n{\nu}
\def\l{\lambda}
\def\L{\Lambda}
\def\t{\tau}
\def\pder#1#2{{\partial #1 \over \partial #2}}
\def\div#1#2{{#1\over #2}}
\def\rout{\ifmmode{r_{\rm out}}\else\hbox{$r_{\rm out}$}\fi}
\def\tmax{\ifmmode{\tau_{\rm max}}\else\hbox{$\tau_{\rm max}$}\fi}
\def\tstd{\ifmmode{\tau_{\rm std}}\else\hbox{$\tau_{\rm std}$}\fi}
\def\vmax{\ifmmode{v_{\rm max}}\else\hbox{$v_{\rm max}$}\fi}
\def\muE{\ifmmode{\mu_{\rm E}}\else\hbox{$\mu_{\rm E}$}\fi} 
\def\pE{\ifmmode{p_{\rm E}}\else\hbox{$p_{\rm E}$}\fi} 
\def\bmax{\ifmmode{\b_{\rm max}}\else\hbox{$\b_{\rm max}$}\fi}
\def\kms{\hbox{$\,$km$\,$s$^{-1}$}}
\def\ergs{\hbox{$\,$erg$\,$s$^{-1}$}}
\def\kpc{\hbox{$\,$kpc} }
\def\ang{\hbox{\AA}}
\def\mag{\hbox{$\rm\,mag$}}
\def\Msun{\hbox{$\,$M$_\odot$} }
\def\Lsun{\hbox{$\,$L$_\odot$} }
\def\Teff{\hbox{$\,T_{\rm eff}$} }
\def\Tcol{\hbox{$\,T_{\rm col}$} }
\def\alog#1{\times 10^{#1}}
\def\rin{\hbox{$r_{\rm in}$} }
\def\rout{\hbox{$r_{\rm out}$} }
\def\pgas{\hbox{$P_{\rm gas}$} }
\def\chistd{\ifmmode{\chi_{\rm std}}\else\hbox{$\chi_{\rm std}$}\fi}

\def\gcm{g~cm$^{-3}$}
\def\is{s$^{-1}$}
\def\k{\,{\rm K}}
\def\K{\,{\rm K}}
\def\msol{$M_\odot$}
\def\foe{10^{51}}
\def\nni{$^{56}$Ni}
\def\xni{{\rm X}_{\rm Ni}}
\def\vno{v_0}
\def\ve{v_e}
\def\teff{{\rm T}_{\rm eff}}
\def\tcirc{{\rm T}_{\rm circ}}

\def\singlet#1{\hbox{$ ^1$#1}}
\def\doublet#1{\hbox{$ ^2$#1}}
\def\triplet#1{\hbox{$ ^3$#1}}
\def\quartet#1{\hbox{$ ^4$#1}}
\def\quintet#1{\hbox{$ ^5$#1}}
\def\sextet#1{ \hbox{$ ^6$#1}}

\def\lstar{\ifmmode{\Lambda^*}\else\hbox{$\Lambda^*$}\fi} 
\def\Rop{\ifmmode{[R_{ij}]}\else\hbox{$[R_{ij}]$}\fi}
\def\Rij{\Rop}
\def\Rji{\ifmmode{[R_{ji}]}\else\hbox{$[R_{ji}]$}\fi}
\def\Rstar{\ifmmode{[R_{ij}^*]}\else\hbox{$[R_{ij}^*]$}\fi}
\def\Rijstar{\Rstar}
\def\Rjistar{\ifmmode{[R_{ji}^*]}\else\hbox{$[R_{ji}^*]$}\fi}
\def\DRji{\ifmmode{[\Delta R_{ji}]}\else\hbox{$[\Delta R_{ji}]$}\fi}
\def\DRij{\ifmmode{[\Delta R_{ij}]}\else\hbox{$[\Delta R_{ij}]$}\fi}

\def\Jb{{\bar J}}
\def\Jnew{{\bar J_{\rm new}}}
\def\Jold{{\bar J_{\rm old}}}
\def\Jfs{{\bar J_{\rm fs}}}
\def\Snew{{S_{\rm new}}}
\def\Sold{{S_{\rm old}}}
\def\Amat{\mat{A}}             

\def\ns{\ifmmode{N_{\rm s}}          
        \else\hbox{$N_{\rm s}$}\fi}

\def\peq{\mathbin{\hbox{$+$}\hbox{$=$}}}

\def\mat#1{{\bf #1}}     
\def\vek#1{{#1}}         

\newcount\eqcount
\eqcount=0
\def
  \nummer{
    \global\advance\eqcount by 1
    (\the\eqcount)
  }

\def
  \numadv{
    \global\advance\eqcount by 1
  }

\def
   \numout#1{
     (\the\eqcount #1)
  }

\def\ivek#1#2{\ifmmode{\vek{I}^{#1}_{#2}}
        \else\hbox{$\vek{I}^{#1}_{#2}$}\fi}

\def\ip#1{\ivek{+}{#1}}      
\def\im#1{\ivek{-}{#1}}      

\def\tmat#1#2{\ifmmode{\mat{t}^{#1}_{#2}}
        \else\hbox{$\mat{t}^{#1}_{#2}$}\fi}
\def\rmat#1#2{\ifmmode{\mat{r}^{#1}_{#2}}
        \else\hbox{$\mat{r}^{#1}_{#2}$}\fi}
\def\bvek#1#2{\ifmmode{\beta^{#1}_{#2}}
        \else\hbox{$\beta^{#1}_{#2}$}\fi}

\def\tpi#1{\tmat{+}{#1}}
\def\tmi#1{\tmat{-}{#1}}
\def\rmi#1{\rmat{-}{#1}}
\def\rpi#1{\rmat{+}{#1}}
\def\bpi#1{\bvek{+}{#1}}
\def\bmi#1{\bvek{-}{#1}}

\def\tp{\tmat{+}{}}          
\def\tm{\tmat{-}{}}          
\def\rmm{\rmat{-}{}}         
\def\rp{\rmat{+}{}}          
\def\bp{\bvek{+}{}}          
\def\bm{\bvek{-}{}}          
\def\tpm{\tmat{\pm}{}}       
\def\rpm{\rmat{\pm}{}}       
\def\bpm{\bvek{\pm}{}}       

\def\lp{\ifmmode{\lambda^+_\tau}           
        \else\hbox{$\lambda^+_\tau$}\fi}
\def\lm{\ifmmode\lambda^-_\tau             
        \else\hbox{$\lambda^-_\tau$}\fi}

\baselineskip=12pt

\bibliographystyle{natbib-apj}

\title{Parallel Implementation of the {\tt PHOENIX} Generalized Stellar
Atmosphere Program. II: Wavelength Parallelization}

\author{
E.~Baron\altaffilmark{1}
and Peter H. Hauschildt\altaffilmark{2}}

\altaffiltext{1}{Dept. of Physics and Astronomy, University of
Oklahoma, 440 W.  Brooks, Rm 131, Norman, OK 73019-0225;
baron@mail.nhn.ou.edu}

\altaffiltext{2}{Dept. of Physics and Astronomy \& Center for
Simulational Physics, University of Georgia, Athens, GA 30602-2451;
yeti@hal.physast.uga.edu}

\accepted{Sept.\ 21, 1997}

\begin{abstract}
We describe an important addition to the parallel implementation of
our generalized NLTE stellar atmosphere and radiative transfer computer
program \phoenix. In a previous paper in this series we described data
and task parallel algorithms we have developed for radiative transfer,
spectral line opacity, and NLTE opacity and rate calculations.
These algorithms divided the work spatially or by spectral lines,
that is distributing the radial zones, individual spectral lines,
or characteristic rays among different processors and employ, in
addition task parallelism for logically independent functions (such as
atomic and molecular line opacities). For finite, monotonic velocity
fields, the radiative transfer equation is an initial value problem in
wavelength, and hence each wavelength point depends upon the previous
one. However, for sophisticated NLTE models of both static and moving
atmospheres needed to accurately describe, e.g., novae and supernovae,
the number of wavelength points is very large (200,000--300,000) and hence
parallelization over wavelength can lead both to considerable speedup
in calculation time and the ability to make use of the aggregate memory
available on massively parallel supercomputers. Here, we describe an
implementation of a pipelined design for the wavelength parallelization
of \phoenix, where the necessary data from the processor working on
a previous wavelength point is sent to the processor working on the
succeeding  wavelength point as soon as it is known.  Our implementation
uses a MIMD design based on a relatively small number of standard \MPI\
library calls and is fully portable between serial and parallel computers.
\end{abstract}

\section{Introduction}

Spectroscopy is one of the most important tools in all of astrophysics. It
is through the use of spectroscopy that we have discovered the
cosmological expansion and determined the elemental composition of the
sun. Currently detailed spectroscopic analyses are used to date the age of
the galaxy \cite[]{cowanetal97}, to determine the structure, energies,
and compositions of novae \cite[]{phhnov95,phhnovfe296,novand86pap}
and supernovae \cite[]{b93j3,b93j4,b94i1,nug1a95,nugseq95,nughydro97}
to probe the conditions at the time of galaxy formation via examining
damped Lyman alpha clouds at high redshift \cite[cf. ][]{pwlya97},
and to confirm the reality of claims for the discovery of sub-stellar
objects \cite[]{araa}.

We have developed the spherically symmetric special
relativistic non-LTE generalized radiative transfer
and stellar atmosphere computer code {\tt PHOENIX}
\cite[]{phhs392,phhcas93,hbfe295,phhnov95,faphh95,phhnovfe296,snefe296}
which can handle very large model atoms as well as line blanketing by
millions of atomic and molecular lines.  This code is designed to be
very flexible, it is used to compute model atmospheres and synthetic
spectra for, e.g., novae, supernovae, M and brown dwarfs, O to M giants,
white dwarfs and accretion disks in Active Galactic Nuclei (AGN); and
it is highly portable.  We include a large number of line transitions
and solve the radiative transfer equation for each of them without using
simple approximations (like the Sobolev approximation), and therefore the
line profiles must be resolved in the co-moving (Lagrangian) frame. This
requires many wavelength points (we typically use 150,000 to 300,000).
Since the CPU time scales linearly with the number of wavelength points,
the CPU time requirements of such a calculation are large. In addition,
NLTE radiative rates for both line and continuum transitions must be
calculated and stored at every spatial grid point for each transition,
which requires large amounts of storage and can cause significant
performance degradation if the corresponding routines are not optimally
coded.

In order to take advantage of the enormous computing power and vast
aggregate memory sizes of modern parallel supercomputers, both potentially
allowing much faster model construction as well as more sophisticated
models, we have developed a parallel version of {\tt PHOENIX}. Since the
code uses a modular design, we have implemented different parallelization
strategies for different modules in order to maximize the total parallel
speed-up of the code. In addition, our implementation allows us to
change the distribution of computational work onto different
nodes both
via input files and dynamically during a model run, which gives a high
degree of flexibility to optimize the performance for both a number of
different parallel supercomputers (we are currently using {\tt IBM SP2}s,
{\tt SGI Origin 2000}s, {\tt HP/Convex SPP-2000}s, and {\tt Cray T3E}s)
and for different model parameters.

Since we have both large CPU and memory requirements we have developed
the parallel version of the code using the \MPI\ message passing library
\cite[]{mpistd}.  The single processor speed of a machine like the \SP2\
is moderately high so that even a small number of additional processors
can lead to significant speed-up. We have chosen to work with the \MPI\
message passing interface, since it is both portable \cite[public domain
implementations of \MPI\ are readily available cf.~][]{mpich}, running on
dedicated parallel machines and heterogeneous workstation clusters and
it is available for both distributed and shared memory architectures.
For our application, the distributed memory model is in fact easier to
use than a shared memory model, since then we do not have to worry about
locks and synchronization, on {\em small} scales and, in addition, we
retain full control over interprocess communication.  This is especially
clear once one realizes that it can be more cost-effective to avoid
costly communication by executing identical code on many processing
elements (or {\em nodes}) as long as the impact on the total CPU time is small, rather
than parallelizing each individual module with the corresponding high
cost of communication and loop overhead.  Distributed massively parallel
supercomputers also typically have more aggregate memory, which enables
them to run much larger simulations than traditional serial 
computers.
Our initial parallelization of the
code \cite[]{hbapara97} was straightforward in that we distributed
the computations among the different modules (task parallelism) and
we were further able to sub-divide some of the modules by utilizing
data parallelism in, e.g, the radial coordinate or individual spectral
lines. Thus, \phoenix\ uses both task and data parallelism {\em at the
same time} in order to optimized performance and allow larger model
calculations.

\section{Equations and Problem Description}

The co-moving frame radiative transfer equation for spherically
symmetric flows can be written 
 as \cite[cf.~][]{found84}:

\vbox{\be
&\quad&\gamma (1+\beta\mu)\frac{\partial\inu}{\partial t} + \gamma (\mu +
\beta) \frac{\partial\inu}{\partial r}\nonumber\\
& +& \frac{\partial}{\partial
\mu}\left\{ \gamma (1-\mu^2)\left[ \frac{1+\beta\mu}{r}
\right.\right.\nonumber\\
&\quad&\left.\left. \quad -\gamma^2(\mu+\beta)
\frac{\partial\beta}{\partial r} -  
\gamma^2(1+\beta\mu) \frac{\partial\beta}{\partial
t}\right] \inu\right\} \nonumber\\
&-&  \frac{\partial}{\partial
\nu}\left\{ \gamma\nu\left[ \frac{\beta(1-\mu^2)}{r}
+\gamma^2\mu(\mu+\beta) \frac{\partial\beta}{\partial r}
\right.\right.\nonumber\\
&\quad& \left.\left.\quad  +
\gamma^2\mu(1+\beta\mu) \frac{\partial\beta}{\partial
t}\right]\inu\right\}\label{fullrte}\\
&+&\gamma\left\{\frac{2\mu+\beta(3-\mu^2)}{r}\right.
\nonumber\\
&\quad&\quad
\left. +\gamma^2(1+\mu^2+2\beta\mu)\frac{\partial\beta}{\partial r} + 
\gamma^2[2\mu + \beta(1+\mu^2)]\frac{\partial\beta}{\partial
t}\right\}\inu \nonumber\\
&\quad& = \eta_\nu - \chi_\nu\inu.\nonumber
\label{RTE}\ee}
\noindent
We set  $c=1$; $\beta$ is the velocity; and
$\gamma = (1-\beta^2)^{-1/2}$ is the usual Lorentz factor. 
Equation~\ref{fullrte} is a integro-differential equation, since the
emissivity $\eta_\nu$ contains \jnu\unskip, the zeroth angular moment of
\inu\unskip:
\[ \eta_\nu = \kappa_\nu \snu + \sigma_\nu \jnu, \]
 and 
\[ \jnu = 1/2 \int_{-1}^{1} d\mu\, \inu, \]
where $\snu$ is the source function, $\kappa_\nu$ is the absorption
opacity, and $\sigma_\nu$ is the scattering opacity.
With
the assumption of time-independence $\frac{\partial\inu}{\partial t} =
0$ and a monotonic velocity field Eq.~\ref{fullrte} becomes a boundary-value problem in
the spatial coordinate and an initial value problem in the frequency
or wavelength coordinate. The equation can be written in operator form
as:
\be
\jnu = \L_\nu \snu,
\ee
where $\L$ is the lambda-operator. 

Implicit in the solution of these equations is obtaining correct
expressions for the opacity and the source function, both of which
depend on the level populations of the material at each spatial
point. Thus, one is forced to include the auxiliary equations which
include the steady state rate equations (transitions into a given
level are balanced by those out of that level), the NLTE equation of
state which enforces charge and mass conservation, and the radiative
equilibrium equation which enforces energy conservation. These
auxiliary equations of course involve the radiation field, which makes
the problem highly non-linear.

\subsection{Definition of terms}

We define a {\em task} as a logical unit of code that treats an aspect
of the physics of the simulation, such as radiative transfer or NLTE
rate calculations. In many cases, different tasks are independent and
can be executed in parallel, we call this coarse grained parallelism
{\em task parallelism}.  Within each task, the opportunity of {\em
data parallelism} may exist, which is a fine grained parallelism, e.g.,
on the level of individual loops in which loop-iteration is independent.

We use the term {\em node} to indicate a single processing element of the
parallel computer which is the smallest possible separate computational
unit of the parallel computer. A node might be a single CPU (e.g.,
an IBM SP2 thin node) or it might be a multi-CPU SMP node (e.g., a
dual Pentium Pro system that is part of a networked cluster used as a
parallel machine). Each node is assumed to have local virtual memory
and a means of communicating with the other nodes in the system. In
addition, the node has access to a global filesystem and, possibly, a
local filesystem as well. These assumptions are fulfilled by basically
all of the currently available parallel supercomputer systems (with the
exception of the Cray T3E which does {\em not} support virtual memory).

A single node can execute a number of tasks, either serial on a single
CPU or in parallel, e.g., on an SMP node of a distributed shared-memory
supercomputer. In addition, data parallelism can also be used across
nodes, or any combination of task and data parallelism can be used
simultaneously.

\subsection{Task and data parallel algorithms in \phoenix}

In a previous paper \cite[][Paper I]{hbapara97} we described our
method for parallelizing three separate modules: (1) The radiative
transfer calculation itself, where we divide up the characteristic
rays among nodes and use an {\tt MPI\_REDUCE} to send the \jnu to all
the radiative transfer and NLTE rate computation tasks; (2) the line
opacity which requires the calculation of about 10,000 Voigt profiles
per wavelength point at each radial grid point, here we split the work
amongst the processors both by radial grid point and by dividing up
the individual lines to be calculated among the processors; and (3) the
NLTE calculations.  The NLTE calculations involve three separate parts:
the calculation of the NLTE opacities, the calculation of the rates at
each wavelength point, and  the solution of the NLTE rate equations. In
Paper I we performed all these parallelizations by distribution of the
radial grid points among the different nodes or by distributing sets of
spectral lines onto different nodes. In addition, to prevent communication
overhead, each task computing the NLTE rates is paired on the same node
with and the corresponding task computing NLTE opacities and emissivities
to reduce communication. The solution of the rate equations parallelizes
trivially with the use of a diagonal rate operator.

In the latest version of our code, {\tt PHOENIX 8.1}, we have incorporated
the additional strategy of distributing each NLTE species (the total
number of ionization stages of a particular element treated in NLTE) on
separate nodes. Since different species have different numbers of levels
treated in NLTE (e.g. Fe~II [singly ionized iron] has 617 NLTE levels,
whereas H~I has 30 levels), care is needed to balance the number of
levels and NLTE transitions treated among the nodes to avoid unnecessary
synchronization problems.

In addition to the data parallelism discussed above, the version of
\phoenix\ described in paper I also uses simultaneous task parallelism
by allocating different tasks to different nodes. This can result in
further speed-up and better scalability but requires a careful analysis
of the workload between different tasks (the workload is also a function
of wavelength, e.g., different number of lines that overlap at each
wavelength point) to obtain optimal load balancing.

\section{Wavelength Parallelization}

The division of labor outlined in the previous section requires
synchronization between the radiative transfer tasks and the NLTE tasks,
since the radiation field and the ``approximate $\Lambda$ operator'' must
be passed between them. In addition, our standard model calculations use
50 radial grid points and as the number of nodes increases, so too does
the communication and loop overhead. We found good speedup up to about 5
nodes for a typical supernova calculation, with the speedup close to the
theoretical maximum.  However, for 5 nodes the communication and loop
overheads begin to become significant and it is not economical to use
more than 10 nodes (depending on the machine and the model calculation,
it might be necessary to use more nodes to fit the data in the memory
available on a single node).

Since the number of wavelength points in a calculation is very large
and the CPU time scales linearly with the number of wavelength points, a
further distribution of labor by wavelength points would potentially lead
to large speedups and to the ability to use very large numbers of nodes
available on massively parallel supercomputers. Thus, we have developed
the concept of wavelength ``clusters'' to distribute a set of wavelength
points (for the solution of the frequency dependent radiative transfer)
onto a different set of nodes, see Fig.~\ref{design}.  In order to achieve
optimal load balance and, more importantly, in order to minimize the
memory requirements, each cluster works on a single wavelength point
at any given time, but it may consist of a {\em number} of ``worker''
nodes where the worker nodes use parallelization methods discussed in
paper I. In order to avoid communication overhead, the workers of each
wavelength cluster are {\em symmetric}:  each corresponding worker on
each wavelength cluster performs identical tasks but on a different
set of wavelengths for each cluster. We thus arrange the total number
of nodes $N$ in a rectangular matrix with $n$ columns and $m$ rows,
where $n$ is the number of clusters and $m$ is the number of workers
for each cluster, such that $N=n * m$.

This design allows us to make use of {\em communicator contexts}, a
concept which is built into {\tt MPI}.  The nodes of a given wavelength
cluster are assigned to a single {\tt MPI\_GROUP} (a vertical column in
Fig.~\ref{design}) so that the $m$ nodes of each cluster form their own
{\tt MPI\_GROUP} and have their own \MPI\ {\em communicator} to pass
messages {\em within} a cluster.  We use the task and data parallelism
introduced in paper I within each individual cluster if the $m$ is larger
than one.  In addition to the $n$ {\tt MPI\_GROUP}s for the $m$ workers of
each of the $n$ clusters, we also use $m$ {\tt MPI\_GROUP}s for the $n$
clusters with the corresponding communicators.  These groups can pass
messages within an individual row of Fig.~\ref{design}, thus allowing
the flow of information between wavelength points, this is important for
the solution of the co-moving frame RTE as discussed below.  The code has
been designed so that the number of wavelength clusters $n$, the number
of workers per wavelength cluster $m$, and the task distribution within
a wavelength cluster is arbitrary and can be specified dynamically at
run time.

For a static model atmosphere, all wavelengths and thus wavelength
clusters are completely independent and execute in parallel with {\em no}
communication or synchronization  along the rows of Fig.~\ref{design}.
The basic pre- and post-processing required are illustrated in the
pseudo-code in Fig.~\ref{phoenix}.  However, in order to parallelize the
spectrum calculations for a model atmosphere with a global velocity
field, such as the expanding atmospheres of novae, supernovae or
stellar winds, we need to take the mathematical character of the RTE
into account. For monotonic velocity fields, the RTE is an initial
value problem in wavelength (with the initial condition at the smallest
wavelength for expanding atmospheres and at the largest wavelength
for contracting atmospheres). This initial value problem must be
discretized fully implicitly to ensure stability.  In the simplest case
of a first order discretization, the solution of the RTE for wavelength
point $i$ depends only on the results of the point $i-1$. In order to
parallelize the spectrum calculations, the wavelength cluster $n_i$
computing the solution for wavelength point $i$ must get the specific
intensities from the cluster $n_{i-1}$ computing the solution for
point $i-1$. This suggests a ``pipeline'' solution to the wavelength
parallelization. Note that only the solution of the RTE is affected by
this, the calculation of the opacities and rates remains independent
between different wavelength clusters and remains fully parallelized. In
this case, the wavelength parallelization works as follows: Each cluster
can independently compute the opacities and start the RT calculations
(hereafter called the {\em pre-processing phase}), it then waits until
it receives the specific intensities for the previous wavelength point,
then it finishes the solution of the RTE and {\em immediately} sends the
results to the wavelength cluster calculating the next wavelength point
(to minimize waiting time, this is done with non-blocking send/receives),
then proceeds to calculate the rates etc.\ (hereafter called the {\em
post-processing phase} and the new opacities for its {\em next} wavelength
point and so on.

The important point in this scheme is that each wavelength cluster can
execute the post-processing phase of its current wavelength point and
pre-processing phase of its next wavelength point {\em independently
and in parallel with all other clusters}. This means that the majority
of the total computational work can be done in parallel, leading to a
substantial reduction in wall-clock time per model.  Ideal load balancing
can be obtained by dynamically allocating wavelength points to wavelength
clusters. This requires only primitive logic with no measurable overhead,
however it requires also communication and an arbitration/synchronization
process to avoid deadlocks.  Typically, the number of clusters $n$
(4-64) is much smaller than the number of wavelength points, $n_{\rm
wl}\approx 300,000$, so that at any given time the work required for each
wavelength point is roughly the same for each cluster (the work changes
as the number of overlapping lines changes, for example). Therefore,
a simple {\em round robin} allocation of wavelength points to clusters
(cluster $i$ calculates wavelength points $i$, $n+i$, $2n+i$ and so
on) can be used which will result in nearly optimal performance if the
condition $n \ll n_{\rm wl}$ is fulfilled.

As an example, we consider the the simple case of 16 nodes (with one CPU
each), distributed to 2 wavelength clusters with 8 worker nodes each,
i.e., $n=2$, $m=8$.  Using the round-robin scheme, clusters 1--2 are
allocated to wavelength points 1--2, respectively. Within in each cluster
the work is divided using the combined task and data parallelism discussed
in paper I where the work topology is identical for each cluster.
All clusters begin immediately by executing the various preprocessing
required. Since cluster 1 begins with wavelength point 1, it sets the
initial condition for the co-moving frame RT.  It then solves the RTE
and {\em immediately} sends the specific intensities off to cluster 2
(which is already working on wavelength point 2) using a non-blocking
\MPI\ send.  Cluster 1 then continues to calculate the rates and various
other post-processing at this wavelength point and then immediately
proceeds the pre-processing phase of its next wavelength point, number
3 in this case. At the same time, cluster 2 has finished calculating
the opacities at its wavelength point number 2, done the preprocessing
for solving the RTE and then must wait until cluster 1 has sent it the
specific intensities for the previous wavelength point. It
can then solve the RTE and immediately send its specific intensities on
to cluster 1.  Since node 1 was busy doing post-processing for wavelength
point 1 and pre-processing for wavelength point 3, it may in fact have
the specific intensities from node 2 just in time when it needs them
to continue with the solution of the RTE for wavelength point 3 and so
minimal waiting may be required and the process proceeds in a round robin
fashion.  Because we employ more than one worker per wavelength cluster
the combined task and data parallel method described in \cite{hbapara97}
is used within the wavelength cluster and since the workers are symmetric,
the sending of data must only be done between identical workers on each
wavelength cluster as depicted in Figure~\ref{design}, thus minimizing
the inter-cluster communication.

This scheme has some predictable properties similar to the performance
results for classical serial vector machines. First, for a very small
number of clusters (e.g., two), the speedup will be small because the
clusters would spend a significant amount of time waiting for the results
from the previous cluster (the ``pipeline'' has no time to fill). Second,
the speedup will level off for a very large number of clusters when the
clusters have to wait because some of the clusters working on previous
wavelength points have not yet finished their RTE solution, thus limiting
the minimum theoretical time for the spectrum calculation to roughly the
time required to solve the RTE for all the wavelength points together
(the ``pipeline'' is completely filled).  This means that there is a
``sweet spot'' for which the speedup to number-of-wavelength-clusters
ratio is optimal. This ratio can be further optimized by using the
optimal number of worker nodes per cluster, thus obtaining an optimal
number of total nodes. The optimum will depend on the model atmosphere
parameters, the speed of each node itself and the communication speed,
as well as the quality of the compilers and libraries. 

 The wavelength parallelization has the drawback that it does not
reduce the memory requirement per node compared to runs with a single
wavelength cluster. Increasing the number of worker nodes per cluster
will decrease the memory requirements per node drastically, however,
so that large runs can use both parallelization methods at the same
time to execute large simulations on nodes with limited memory. On a
shared-memory machine with distributed physical memory (such as the
\Origin), this scheme can also be used to minimize memory access latency.

\section{Results of performance tests}

\subsection{Static atmospheres}

In the case of a static model atmosphere ($\beta=0$), Eq.~\ref{RTE} can be
solved independently for each wavelength point because the non-coherent
scattering is handled in the rate-operator formalism. This means that
for static atmospheres the parallelization over wavelength is simple
and involves no communication or synchronization during the spectrum
calculations. For our large set of wavelengths points, this will lead
to good parallel performance. This is illustrated in Fig.~\ref{static}
for a NLTE model atmosphere run with parameters appropriate for the A0V
star Vega ($\Teff=9600\K$, $\logg=4.0$, solar abundances). The model
includes about 4500 NLTE levels with nearly 51000 primary NLTE lines
(with with detailed Voigt profiles for nearly 39000 of them), about
320000 background LTE lines and 340000 secondary NLTE lines (dynamically
selected). The calculation was performed on a grid of about 270000
wavelength points. This is a typical case of a main sequence star NLTE
model. The memory requirements of this calculations are high, therefore,
we had to use at least 2 worker nodes per wavelength cluster on one of
the \IBMSP s that we were using for this test. This model is a static
atmosphere, so that different wavelength points are independent from
each other and no communication between clusters is required until the
spectrum calculation is complete.  Therefore, the scalability of the
calculation is excellent, in particular on the \Origin\ and the \IBMSP\
runs with a single worker node per wavelength cluster. Clearly, it is more
effective for this model type to use the minimum number of worker nodes
per wavelength cluster to minimize communication and other overheads. The
overhead due to a limited number of IO nodes and limited IO bandwidth
available on the production \IBMSP\ we used for the tests reduces the
speedup for large number of nodes when nodes start to compete for the
available IO bandwidth.

Clearly, for a very small number of processors the wavelength
parallelization is less effective than is the spatial
parallelization. This is caused  by processors competing for
IO bandwidth, rather than synchronization problems.  However,
once the number of processors begins to increase, the wavelength
parallelization clearly scales significantly better than does the spatial
parallelization. Therefore, it is optimal to use the minimal number of
worker nodes per wavelength cluster (defined such that the code fits
completely into the memory available at each node) and use as many
wavelength clusters as possible.

\subsection{Expanding atmospheres}

In Fig.~\ref{nova} we show the performance results for a nova model
atmosphere calculation ($\Teff=15000\K$, $L=200000\Lsun$, $\vmax=2000\kms$, solar
abundances) using various configurations running on the same \IBMSP.
The model includes 1775 NLTE levels with 32056 primary NLTE lines, about
1.3 million background LTE lines and about 90000 secondary NLTE lines
(dynamically selected). The calculation was performed on a grid of about
175000 wavelength points. This model is somewhat smaller than our typical
nova models, it was used because it is small enough to run in serial
mode on the \IBMSP\ that we used for the tests.  The behavior of the
parallel performance and scalability is essentially as expected. For a
small number of nodes, the speedup obtained by using wavelength cluster
is smaller than the speedup obtained by using one wavelength cluster
but several worker nodes.  As the number of nodes increases, it is more
effective to use more wavelength clusters than more workers. However,
as the number of clusters increases over a limit (about 8 clusters in
this model), the speedup remains constant if the number of clusters is
increased (and the number of workers remains constant). The optimum load
distribution is thus a combination of all parallelization methods,
depending not only on the machine but also on the workload distribution
of the model calculation itself.

For a very large supernova calculation, we examine both the
scaling and performance tradeoff of spatial versus wavelength
parallelization. Figure~\ref{sn} presents the results of our timing
tests for one iteration of a Type Ic supernova model
atmosphere, with a model temperature $\Tmod = 12,000$~K (the observed
luminosity is given by $L=4\pi R^2 \Tmod^4$), characteristic velocity
$\vno=10000$~\kms, 4666 NLTE levels, 163812 NLTE Gauss lines, 211680
LTE Gauss lines, non-homogeneous abundances, and 260630 wavelength
points. This is among the largest calculations we run and hence it has
the highest potential for synchronization, I/O waiting, and swapping to
reduce performance. It is however, characteristic of the level of detail
needed to accurately model supernovae. This calculation has also been
designed to barely fit into the memory of a single node. The behavior of
the speedup is very similar to the results reported for the nova
test case. The fact that the turnover is at lower number of processor
elements is almost certainly due to the higher I/O and memory
bandwidth required by the larger calculation that the supernova
represents over the nova calculation.

The ``saturation point'' at which the wavelength pipeline fills and no
further speedup can be obtained if more wavelength clusters are used
lies for the machines used here at about 5 to 8 clusters. More clusters
will not lead to larger speedups, as expected. Larger speedups can be
obtained by using more worker nodes per cluster, which also drastically
reduces the amount of memory required on each node.

\section{Summary and Conclusions}

We have discussed the methods that we have implemented in \phoenix\
to parallelize the calculation of wavelength dependent spectra (for
both spectral synthesis and model atmosphere generation). While the
algorithms are simple in the case of static stellar atmospheres,
for moving atmospheres, e.g., the expanding atmospheres of novae
and supernovae or stellar winds, the radiative transfer equation is
coupled between different wavelengths. Therefore, we have developed a
``pipelined'' approach that is used in expanding atmosphere models to
parallelize the spectrum calculation. Combined with the ``spatial''
and ``line'' data and task parallelization reported in paper I, this
new parallelization option can dramatically increase the speed of very
detailed and sophisticated NLTE and LTE stellar atmosphere calculation
with \phoenix.  The parallelization has become a standard feature of
the production version of \phoenix\ and we are routinely using all 3
parallelization options simultaneously to calculate model atmospheres
for a large variety of objects from Brown and M dwarfs to novae and
supernovae on parallel supercomputers. This has drastically increased
our productivity with a comparatively small time and coding investment.
It also forms the basis to much larger calculations that will be required
to appropriately analyze the much improved data that can be expected
from future ground- and space-based observatories.

Our wavelength parallelization combines the methods described in paper
I by combining a number of worker nodes (which employ the task and data
parallel algorithms discussed in paper I) into symmetric ``wavelength
clusters'' which work on different wavelength and that communicate
results (if necessary) between them. This scheme is relatively simple
to implement using the \MPI\ standard and can be used on all parallel
computers, both distributed and shared-memory systems (including clusters
of workstations). It has the advantage of minimizing communication and
it allows us to tailor the code's memory usage to the memory available
on each individual node.

The behavior of the wavelength parallelization can be understood easily
and the speedups are as expected. The parallel scalability of \phoenix\
is comparable to or even better than that of many commercially available
scientific applications. The potential of parallel computing for stellar
atmosphere modeling is enormous, both in terms of problem size and speed
to model construction.  The aggregate memory and computing power of
parallel supercomputers can be used to create extremely detailed models
that are impossible to calculate on vector supercomputers or workstations.

\begin{small}
\noindent{\em Acknowledgments:}
We thank the referee, John Castor, for helpful comments that helped to
greatly improve the manuscript. We also thank David Lowenthal for helpful
discussions on parallel computing. This work was supported in part by
NASA ATP grant NAG 5-3018 and LTSA grant NAG 5-3619 to the University
of Georgia, and by NSF grant AST-9417242, NASA grant NAG5-3505 and an
IBM SUR grant to the University of Oklahoma.  Some of the calculations
presented in this paper were performed on the IBM SP2 of the UGA UCNS,
at the San Diego Supercomputer Center (SDSC), the Cornell Theory Center
(CTC), and at the National Center for Supercomputing Applications (NCSA),
with support from the National Science Foundation, and at the NERSC with
support from the DoE. We thank all these institutions for a generous
allocation of computer time.
\end{small}

\clearpage
\bibliography{refs,sn1a,stars,gals,rte,crossrefs,yeti}

\clearpage

\begin{figure}
\caption{\label{design} The basic design of our parallelization
method, groups of processors are divided up into wavelength clusters
which will work on individual wavelength points, the wavelength
clusters are further divided into worker nodes, where each worker node
is assign a set of specific (e.g., spatially distributed) tasks. Our design
requires that each worker node on all wavelength clusters work on
exactly the same set of tasks, although additional inherently serial
operations can be assigned to one particular master worker, or master
wavelength cluster. This method reduces communication between clusters
to its absolute minimum and allows the maximum speedup.}
\end{figure}

\clearpage

\begin{figure}[t]
\caption[]{\label{phoenix} Pseudo-code for the global layout of
\phoenix. The processing that is required before and after the
radiative transfer is indicated. Both pre- and post-processing phases
can be executed in parallel and independently for all clusters.}
\end{figure}

\begin{figure}[t]
\caption{\label{static} Scalability of the static Vega model atmosphere
test run as function of the number of nodes (processing elements or
nodes) used.  The y-axis gives the speedup obtained relative to the serial
run. The different symbols show the results for different numbers of
worker tasks for each wavelength cluster.}
\end{figure}

\begin{figure}[t]
\caption{\label{nova} Scalability of the Nova model atmosphere test run
as function of the number of nodes (processing elements or nodes) used.
The y-axis gives the speedup obtained relative to the serial run. The
different symbols show the results for different numbers of worker tasks
for each wavelength cluster.}
\end{figure}

\begin{figure}[t]
\caption{\label{sn} Scalability of the Supernova model atmosphere test run
as function of the number of nodes (processing elements or nodes) used.
The y-axis gives the speedup obtained relative to the serial run. The
different symbols show the results for different numbers of worker tasks
for each wavelength cluster.}
\end{figure}

\end{document}